# Magnesium Contact Ions Stabilize the Tertiary Structure of Transfer RNA: Electrostatics Mapped by Two-Dimensional Infrared Spectra and Theoretical Simulations

Jakob Schauss, Achintya Kundu, Benjamin P. Fingerhut, and Thomas Elsaesser*

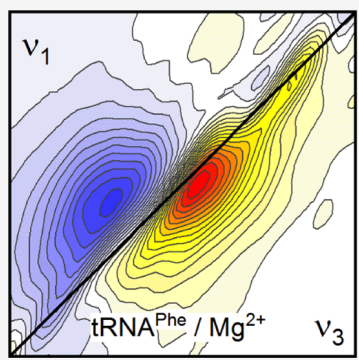
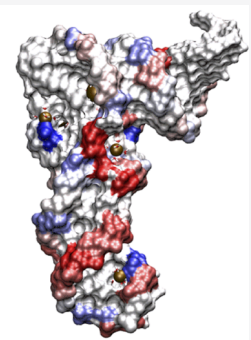

**ABSTRACT:** Ions interacting with hydrated RNA play a central role in defining its secondary and tertiary structure. While spatial arrangements of ions, water molecules, and phosphate groups have been inferred from X-ray studies, the role of electrostatic and other noncovalent interactions in stabilizing compact folded RNA structures is not fully understood at the molecular level. Here, we demonstrate that contact ion pairs of magnesium ($Mg^{2+}$) and phosphate groups embedded in local water shells stabilize the tertiary equilibrium structure of transfer RNA (tRNA). Employing dialyzed tRNA$^{Phe}$ from yeast and tRNA from *Escherichia coli*, we follow the population of $Mg^{2+}$ sites close to phosphate groups of the ribose-phosphodiester backbone step by step, combining linear and nonlinear infrared spectroscopy of phosphate vibrations with molecular dynamics simulations and ab initio vibrational frequency calculations. The formation of up to six $Mg^{2+}$/phosphate contact pairs per tRNA and local field-induced reorientations of water molecules balance the phosphate−phosphate repulsion in nonhelical parts of tRNA, thus stabilizing the folded structure electrostatically. Such geometries display limited sub-picosecond fluctuations in the arrangement of water molecules and ion residence times longer than 1 μs. At higher $Mg^{2+}$ excess, the number of contact ion pairs per tRNA saturates around 6 and weakly interacting ions prevail. Our results suggest a predominance of contact ion pairs over long-range coupling of the ion atmosphere and the biomolecule in defining and stabilizing the tertiary structure of tRNA.

## 1. INTRODUCTION

Electrostatic interactions play a determining role for the secondary and tertiary structures of RNA in the native aqueous environment. The formation of stable macromolecular conformers requires balance of repulsive and attractive electric interactions between the charged and/or polar constituents of the macromolecular structure and its surroundings.[1−3] In particular, repulsive interactions between the negatively charged phosphate groups in the ribose-phosphodiester backbone need to be compensated by positively charged ions and water molecules. Effective shielding of the high negative charge density of RNA is essential for stabilizing the equilibrium structures in their hydration shell and minimizing their total free energy.

Spatial arrangements of positively charged (counter)ions around hydrated RNA, that is, the cations $Na^+$, $K^+$, $Ca^{2+}$, $Mn^{2+}$, or $Mg^{2+}$, have been a subject of calculations based on macroscopic polyelectrolyte theory,[4,5] the (nonlinear) Poisson−Boltzmann (PB) equation,[6,7] and molecular dynamics (MD) simulations, which include the electrostatic interaction potential at the molecular level.[8−11] Such treatments predict a pronounced spatial gradient of the cation concentration, induced by the attractive electrostatic interaction with the negatively charged backbone. This (counter)ion condensation leads to a comparably high ionic concentration close to the backbone, which decreases with radial distance on a length scale of typically 20 Å. Some 70% of positive ions reside within the first 5−6 water layers around the biomolecule. Results of small-angle X-ray scattering studies of short DNA double strands are in qualitative agreement with calculated ion density profiles, without, however, characterizing specific ion sites and/or hydration geometries.[12]

There is a variety of molecular geometries in which the ion ensemble and the embedding water shell interact with the high negative charge density of the RNA backbone. Contact geometries, that is, cations in touch with one or several phosphate groups of the backbone, are characterized by a particularly strong attractive interaction at the expense of a partial desolvation of the ion. Comparably long ion residence times up to microseconds have been reported.[13,14] Contact geometries have been identified in high-resolution X-ray diffraction studies of RNA[15,16] and characterized dynamically by vibrational spectroscopy and theoretical analysis of model systems.[17,18] In contrast, ions separated by several water layers experience a substantially weaker interaction with the backbone. They are part of the so-called "diffuse ion atmosphere" and maintain a diffusive mobility. Because of thermally induced fluctuations in ion position, the ion atmosphere has remained elusive in X-ray diffraction but nevertheless exerts a fluctuating electric force on the hydrated RNA structure.

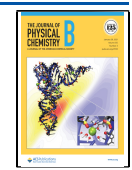

In addition to the cations, the dipolar water molecules of the hydration shell make a significant contribution to the overall electrostatic potential.[19,20] They represent sources of electric fields with a strength of up to 100 MV/cm and simultaneously screen attractive and repulsive electrostatic interactions. The positions and orientations of water molecules adapt to the total Coulomb force they experience. At the same time, water molecules undergo fluctuations on time scales between 50 fs and several tens of picoseconds. The role of this complex molecular ensemble for stabilizing secondary and tertiary structures of RNA as well as the relevant many-body interactions are not understood at the molecular level. Even the spatial range of electric forces and the local hydration geometries of RNA in the presence of ions are barely characterized. Such issues call for experimental probes which map specific interaction sites and their local dynamics.

In this article, we address the fundamental role of $Mg^{2+}$ ions for stabilizing the prototypical equilibrium structure of transfer RNA (tRNA), a central player in translation steps of protein synthesis. Depending on the specific species, tRNA contains 75–90 nucleotides arranged in a folded cloverleaf structure.[16,21,22] The structure of phenylalanine tRNA from yeast (tRNA$^{Phe}$) has been determined by X-ray diffraction with a high spatial resolution of better than 2 Å (ref 16) and is shown in Figure 1a. It consists of the acceptor stem, the $T\Psi C$ loop, the D loop, the variable loop, and the anticodon loop. The acceptor stem contains a single-strand 3′-end (top right of Figure 1a) which protrudes from its double-strand part and serves for attaching the amino acid phenylalanine for protein synthesis. The anticodon loop contains the specific base sequence (O2′-methyl-guanosine–adenosine–adenosine) for reading out complementary messenger RNA. The different loops are connected to double-strand stem regions with paired nucleobases. Other tRNA structures differ in the total number and sequence of standard and nonstandard nucleobases and preserve a folded cloverleaf tertiary structure.[23]

X-ray diffraction with a spatial resolution better than 2 Å (ref 16) has identified 11 binding sites of divalent metal ions, the majority of which are $Mg^{2+}$ ions close to positions of phosphate groups in the folded backbone. Of particular interest are the positions M1, M3, and M7 (Figure 1a) in vicinity of the D-loop where the bending of the backbone results in a separation of neighboring $(PO_2)^-$ oxygens below 4 Å, that is, substantially less than in the A-helical parts of the structure. X-ray diffraction data suggest that such sites are preferentially populated by $Mg^{2+}$ ions with the $(PO_2)^-$ oxygens being part of the first solvation shell of the ion or separated by a single water layer. Such assignments have been challenged by findings of the nonlinear PB model where binding of $Mg^{2+}$ to yeast tRNA$^{Phe}$ has been interpreted on the basis of a single class of ions that retain a complete water shell and stabilize the RNA structure by long-range electrostatic interactions.[24]

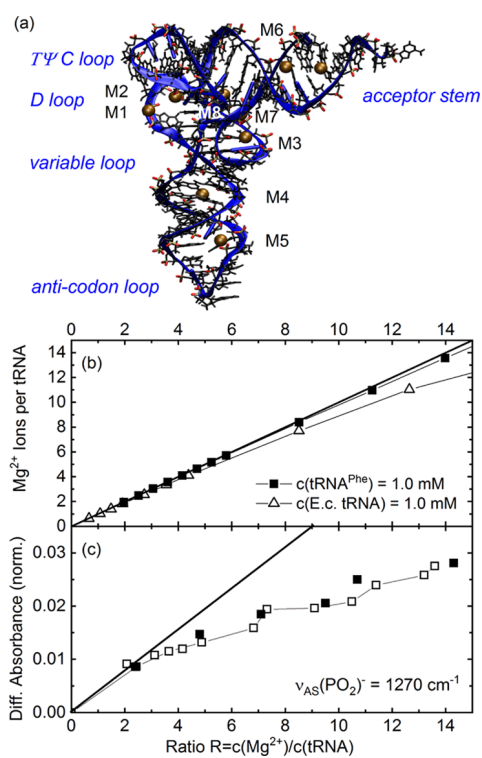

**Figure 1.** (a) Molecular structure of tRNA$^{Phe}$ with the different structural subunits as indicated. The symbols M1 to M8 mark the position of $Mg^{2+}$ ions observed in MD simulations, in good agreement with positions derived from high-resolution X-ray diffraction patterns.[16] (b) Magnesium binding by dialyzed tRNA$^{Phe}$ (solid squares) and E.c. tRNA (open triangles) as determined by fluorescence titration measurements. The average number of $Mg^{2+}$ ions bound per tRNA is plotted as a function of $R = c(Mg^{2+})/c(tRNA)$, the total $Mg^{2+}$ concentration $c(Mg^{2+})$ in units of the concentration $c(tRNA)$ of tRNA in water. The thick solid line corresponds to a scenario in which all $Mg^{2+}$ ions are bound to tRNA. (c) Differential phosphate stretching infrared absorbance of tRNA$^{Phe}$ (solid squares) and E.c. tRNA (open squares) in the presence of $Mg^{2+}$ ions. The differential absorbance $\Delta A = [A(c(Mg^{2+})) - A_0]/A_{ref}$ at 1270 cm$^{-1}$ (cf. Figure 2b) is plotted vs $R$ [$A(c(Mg^{2+}))$: absorbance with $Mg^{2+}$ excess; $A_0$: absorbance of tRNA in water; $A_{ref}$: peak absorbance at 1240 cm$^{-1}$ for $R = 0$]. The absorption around 1270 cm$^{-1}$ is dominated by $Mg^{2+}$–phosphate CIPs.

In our experiments, we study local interactions between $Mg^{2+}$ ions and phosphate groups in the backbones of tRNA$^{Phe}$ and, for comparison, tRNA from *Escherichia coli* (E.c. tRNA). tRNA$^{Phe}$ is chosen because of its well-characterized structure which serves as the starting point for theoretical modeling of electrostatics at the molecular level. E.c. tRNA represents a mixture of different tRNA structures and serves for benchmarking the tRNA$^{Phe}$ results in a wider range of structures. In all systems, local interactions are directly probed via their impact on vibrations of the phosphate groups. Asymmetric $(PO_2)^-$ stretching vibrations $\nu_{AS}(PO_2)^-$ thus serve as sensitive noninvasive probes of the local interaction potentials and allow to map the local dynamics. Samples of dialyzed tRNA$^{Phe}$ and dialyzed E.c. tRNA with a specific $Mg^{2+}$ concentration in a range from zero to approximately 15 $Mg^{2+}$ ions per tRNA are employed to follow the formation of contact geometries step-by-step with the help of linear and femtosecond nonlinear infrared spectroscopy. Such results are analyzed with the help of extensive theoretical calculations, including MD simulations of the molecular ensemble up to microsecond simulation times and ab-initio simulations of the asymmetric $(PO_2)^-$ stretching vibrations $\nu_{AS}(PO_2)^-$.

## 2. MATERIALS AND METHODS

Dialyzed tRNA samples are prepared with a defined $Mg^{2+}$ content and characterized by fluorescence titration and vibrational spectroscopy. The formation of contact ion pairs with tRNA phosphate groups is followed by linear and nonlinear 2D infrared spectroscopy of the asymmetric phosphate stretching vibration and analyzed by quantum mechanics/molecular mechanics (QM/MM) calculations. A detailed description of materials and methods is given in Supporting Information.

## 3. RESULTS

The tRNA samples are prepared in an aqueous buffer solution with tRNA$^{Phe}$ and E.c. tRNA, respectively (both from Aldrich). The E.c. tRNA sample represents a mixture of tRNAs with different anticodon units and amino acid acceptor arms (cf. Figure 1a). The samples of millimolar tRNA concentration are repeatedly dialyzed by following the procedures described in ref 25. In this way, the magnesium content in the tRNA sample is reduced to less than one $Mg^{2+}$ ion per tRNA entity on average. To this aqueous reference solution, defined amounts of a solution of MgCl$_2$ in water are added in order to generate a well-defined content of $Mg^{2+}$ ions in the sample.

The $Mg^{2+}$ ions interacting with tRNA$^{Phe}$ and E.c. tRNA need to be distinguished from the $Mg^{2+}$ ions fully solvated in the water environment. To this end, the fraction of $Mg^{2+}$ ions interacting with the tRNAs is determined with the help of the fluorescence titration method outlined in refs 25,26 (Figure 1b) and compared to results from infrared spectroscopy (Figure 1c). Figure 1b (symbols) displays the average number of interacting $Mg^{2+}$ ions per tRNA$^{Phe}$ (solid squares) and E.c. tRNA (open triangles) as a function of $R = c(Mg^{2+})/c(tRNA)$, the ratio of the total $Mg^{2+}$ and the tRNA concentrations. The tRNA concentration is 1 mM in both cases. The number of interacting $Mg^{2+}$ ions rises linearly with increasing concentration ratio up to $R \approx 7$. In this range, practically all added $Mg^{2+}$ ions interact with the tRNAs, as is evident from the comparison with the reference line (black solid line). At higher $Mg^{2+}$ concentration, the fraction of interacting $Mg^{2+}$ displays a weaker rise and eventually saturates (not shown, cf. ref 25). Variations of the millimolar tRNA concentrations by a factor of 2–3 have a negligible impact on this behavior.

Linear infrared absorption spectra in the range of the asymmetric $(PO_2)^-$ stretching vibrations $\nu_{AS}(PO_2)^-$ of the tRNA$^{Phe}$ backbone are summarized in Figure 2. Figure 2a shows the infrared bands consisting of two strong components with maxima at 1220 and 1241 cm$^{-1}$ and a shoulder around 1270 cm$^{-1}$. Upon addition of $Mg^{2+}$ ions, the infrared absorption undergoes systematic changes, that is, a decrease of absorption on the two strong components and an increase of absorption between 1250 and 1300 cm$^{-1}$. To display this behavior more clearly, the absorbance difference of the $Mg^{2+}$-containing samples and the sample without $Mg^{2+}$ content was calculated. The resulting spectra in Figure 2b clearly exhibit a differential absorption band around 1270 cm$^{-1}$, which rises proportional to the $Mg^{2+}$ concentration with minor changes of line shape. The vibrational spectra of E.c. tRNA behave in a very similar way (not shown). As will be discussed in detail below, the absorption band at 1270 cm$^{-1}$ is induced by the formation of contact ion pairs (CIPs) of $Mg^{2+}$ ions with phosphate groups.

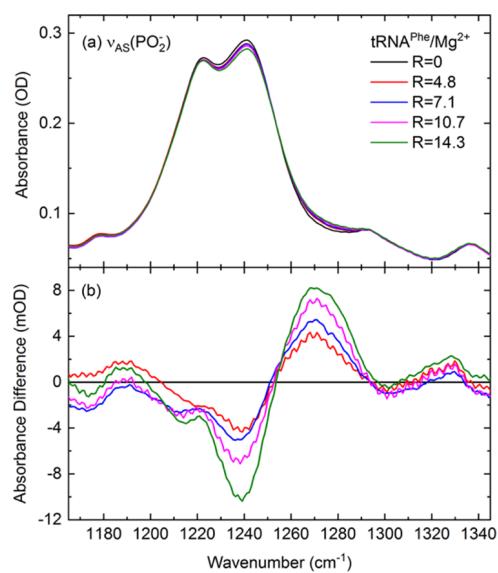

Figure 2. Linear infrared absorption spectra of asymmetric phosphate stretching vibrations $\nu_{AS}(PO_2)^-$ of dialyzed tRNA$^{Phe}$ in water. (a) Infrared absorbance A is plotted as a function of wavenumber for a sample without $Mg^{2+}$ ions ($c(tRNA^{Phe}) = 4.2$ mM) and for different $Mg^{2+}$ concentrations $c(Mg^{2+})$ (colored solid lines). The quantity $R = c(Mg^{2+})/c(tRNA^{Phe})$ is the ratio of $Mg^{2+}$ to tRNA$^{Phe}$ concentration. (b) Differential absorbance spectra $\Delta A = A(c(Mg^{2+})) - A_0$ of tRNA$^{Phe}$ for different magnesium excess concentrations ($A(c(Mg^{2+}))$: absorbance with $Mg^{2+}$ excess; $A_0$: absorbance without $Mg^{2+}$ ions). The rise of absorption around 1270 cm$^{-1}$ is a hallmark of $Mg^{2+}$–phosphate contact ion pair (CIP) formation.

In Figure 1c, the peak value of differential absorbance at 1270 cm$^{-1}$ normalized to the peak absorbance $A_{ref}$ at 1240 cm$^{-1}$ for $R = 0$ is plotted as a function of the ratio $R$ of total $Mg^{2+}$ to tRNA concentration for both tRNA$^{Phe}$ (solid squares) and E.c. tRNA (open squares). Normalization to $A_{ref}$ makes data comparable which were taken with slightly different tRNA concentrations and sample thicknesses. In the range from $R = 0$ to 2, the differential absorbance increases by some 0.01. The linear extrapolation of this absorbance increase to higher ratios $R$ is shown as thick black line. However, the experimental values for both tRNA$^{Phe}$ and E.c. tRNA (symbols) display a more gradual rise which is much weaker than the increase of interacting $Mg^{2+}$ ions plotted in Figure 1b. This discrepancy shows that only a fraction of $Mg^{2+}$ ions interacting with tRNA contribute to this particular absorption band, that is, are accommodated as CIPs with tRNA phosphate groups.

The measurements of linear infrared absorption spectra were complemented by extensive two-dimensional infrared (2D-IR) experiments in order to separate and characterize the different types of $\nu_{AS}(PO_2)^-$ excitations, including their ultrafast dynamics, in depth. Figure 3 summarizes 2D-IR spectra for (a–e) dialyzed tRNA$^{Phe}$ at different concentration ratios $R = c(Mg^{2+})/c(tRNA^{Phe})$ and (f) E.c. tRNA for $R = c(Mg^{2+})/c(E.c. tRNA) = 15$. The absorptive 2D signal given as the real part of the sum of the rephasing and non-rephasing signal is shown as a function of excitation frequency $\nu_1$ (ordinate) and detection frequency $\nu_3$ (abscissa). The yellow-red contours represent the 2D signals on the v = 0 to 1 transitions of the different vibrations, caused by bleaching of the v = 0 ground state and stimulated emission from the v = 1 state. The blue contours are 2D signals on the v = 1 to 2 transitions of excited oscillators.

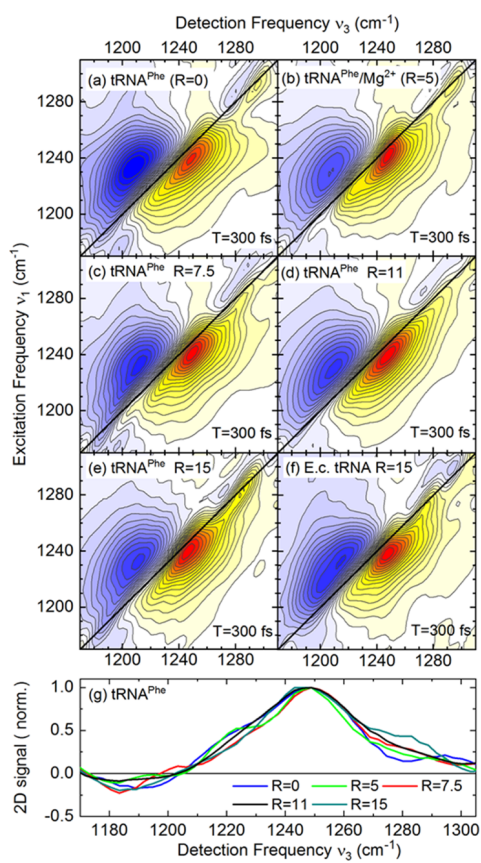

**Figure 3.** 2D-IR spectra of tRNA in water in the range of the asymmetric phosphate stretching ($\nu_{AS}(PO_2)^-$) band. (a−e) 2D-IR spectra of dialyzed tRNA$^{Phe}$ for increasing $Mg^{2+}$ concentration $c(Mg^{2+})$. The quantity $R = c(Mg^{2+})/c(tRNA)$ is the ratio of $Mg^{2+}$ to tRNA concentration. The absorptive 2D signal is plotted as a function of the excitation frequency $\nu_1$ and the detection frequency $\nu_3$. Yellow-red contours represent signals due to the fundamental (v = 0 to 1) transition, and blue contours, the excited state v = 1 to 2 absorption. The signal change between neighboring contour lines is 7.5%. With increasing $Mg^{2+}$ content, there is a pronounced increase of the 2D-IR signal around 1270 cm$^{-1}$. (f) 2D-IR spectrum of E.c. tRNA for R = 15. (g) Cuts of the 2D-IR spectra of tRNA$^{Phe}$ along a diagonal line through ($\nu_1$, $\nu_3$) = (1240, 1250) cm$^{-1}$.

Three components around 1220, 1245, and 1270 cm$^{-1}$ are clearly discerned in the v = 0 to 1 2D signals and the cuts of the tRNA$^{Phe}$ spectra along a diagonal line running parallel to $\nu_3$ = $\nu_1$ through the maximum of the 2D signal at ∼1250 cm$^{-1}$ (Figure 3g). Compared to the linear absorption spectra (Figure 2a), the relative amplitudes of the three components markedly changed, with a pronounced enhancement of the contribution around 1270 cm$^{-1}$. The origin of this behavior will be discussed below. All line shapes are elongated along the diagonal, a fact reflecting inhomogeneous broadening due to a distribution of vibrational frequencies of phosphate groups with a different local environment. Cuts of the 2D spectra along the antidiagonal direction are presented in Supporting Information and reveal a smaller antidiagonal width of the 2D signal contours around 1270 cm$^{-1}$ than of those at lower-detection frequencies. The 2D-IR spectra of E.c. tRNA at R = 15 and at lower values of R (not shown) display a very similar behavior.

Results of femtosecond pump−probe experiments with tRNA$^{Phe}$ are presented in Supporting Information. In the absence of $Mg^{2+}$ ions, such measurements give lifetimes of the v = 1 state of the $\nu_{AS}(PO_2)^-$ vibrations of 290 ± 30 fs, similar to decay times observed with other DNA and RNA structures.[19,27,28] Upon addition of $Mg^{2+}$, one observes a slowing down of the overall decay at probe frequencies above 1260 cm$^{-1}$. This behavior is accounted for by a biexponential signal decay with time constants of 290 and 700 fs (cf. Supporting Information).

The experiments were complemented by atomistic MD simulations extending into the microsecond timescale and extensive mixed QM/MM simulations of $\nu_{AS}(PO_2)^-$ vibrations of yeast tRNA$^{Phe}$. The QM/MM simulations of tRNA$^{Phe}$ backbone vibrations reveal a distribution of $\nu_{AS}(PO_2)^-$ frequency positions due to different local hydration geometries of the $(PO_2)^-$ groups and specific ion interactions at different positions of the tRNA$^{Phe}$ surface. High accuracy in simulations of tRNA$^{Phe}$ backbone vibrations is obtained by treating the sugar−phosphate backbone together with water molecules in the first solvation shell of the $(PO_2)^-$ groups and water molecules in the first solvation shell of CIPs on the QM level of theory. Specifically, QM/MM models for the evaluation of the vibrational frequencies of the sugar−phosphate backbone of tRNA$^{Phe}$ (Figures 4a,b and S9) were constructed by considering two adjacent phosphate groups and the three bridging ribose moieties in the QM region. Additionally, the first solvation shell of phosphate groups was considered in the QM region containing first-shell water molecules, contact ions, and waters in the first solvation shell of the ions. The QM region of vibrational frequency simulations comprises, depending on the particular hydration geometry, 52 sugar-phosphate backbone atoms, 7−18 water molecules, and 0−2 ions (73−108 QM atoms, 682−929 atomic basis functions; see Supporting Information). Figure 4a compares the simulated linear infrared absorption spectrum of tRNA$^{Phe}$ in the frequency range of the $\nu_{AS}(PO_2)^-$ vibrations to the experimental spectrum of undialyzed tRNA$^{Phe}$ in water. We find excellent agreement in frequency position of $\nu_{AS}(PO_2)^-$ covering a range from ∼1180 to 1290 cm$^{-1}$ while some deviation in the intensity in the different frequency ranges is recognized.

A characteristic feature of the experimental linear infrared absorption spectra of both tRNA$^{Phe}$ and E.c. tRNA is the increase in absorption between 1250 and 1300 cm$^{-1}$ upon addition of $Mg^{2+}$ ions. To characterize the molecular geometries of tRNA$^{Phe}$ that contribute to this spectral range, we have analyzed the contributions of CIPs (blue lines), solvent-shared ion pairs (SSIPs) of $(PO_2)^-$ groups with $Mg^{2+}$ ions (red lines), and all other $(PO_2)^-$ groups (black) to the vibrational density of states (DOS, inset Figure 4a). We find a predominant contribution from CIPs in the frequency range $\nu_{AS}(PO_2)^-$ = 1247−1285 cm$^{-1}$, mimicking the experimental observation. A blue-shift of vibrational frequency requires the integration of one of the $(PO_2)^-$ oxygens in the essentially octahedral first solvation layer around the $Mg^{2+}$ ion, similar to what has been observed in model systems.[17,18,28] Because of the short $Mg^{2+}$−oxygen distance of approximately 2.1 Å, the vibrational excitation probes the repulsive part of the interaction potential and, thus, a blue-shift arises. There is a single CIP with absorption at a much lower frequency $\nu_{AS}(PO_2)^-$ = 1219 cm$^{-1}$. The lower $\nu_{AS}(PO_2)^-$ frequency is due the particular geometric structure of the CIP being subject to steric constraints of the tRNA$^{Phe}$ sugar–phosphate backbone. While direct $Mg^{2+}$ coordination to one of the $(PO_2)^-$ oxygens is preserved ($Mg^{2+}$–$(PO_2)^-$ oxygen distance < 2.2 Å), the angular arrangement of the $(PO_2)^-$ group and the ion is different compared to $(PO_2)^-$ units with a blue-shifted $\nu_{AS}(PO_2)^-$ frequency. The bent arrangement of the CIP ($Mg^{2+}$⋯O1P⋯P angle ∼ 131°) reduces the impact of the repulsive part of the interaction potential at a short $Mg^{2+}$–$(PO_2)^-$ oxygen distance, and the blue-shift is diminished.[18,28] For SSIPs, a moderate red-shift of $\nu_{AS}(PO_2)^-$ to the frequency range 1150–1247 cm$^{-1}$ is found.

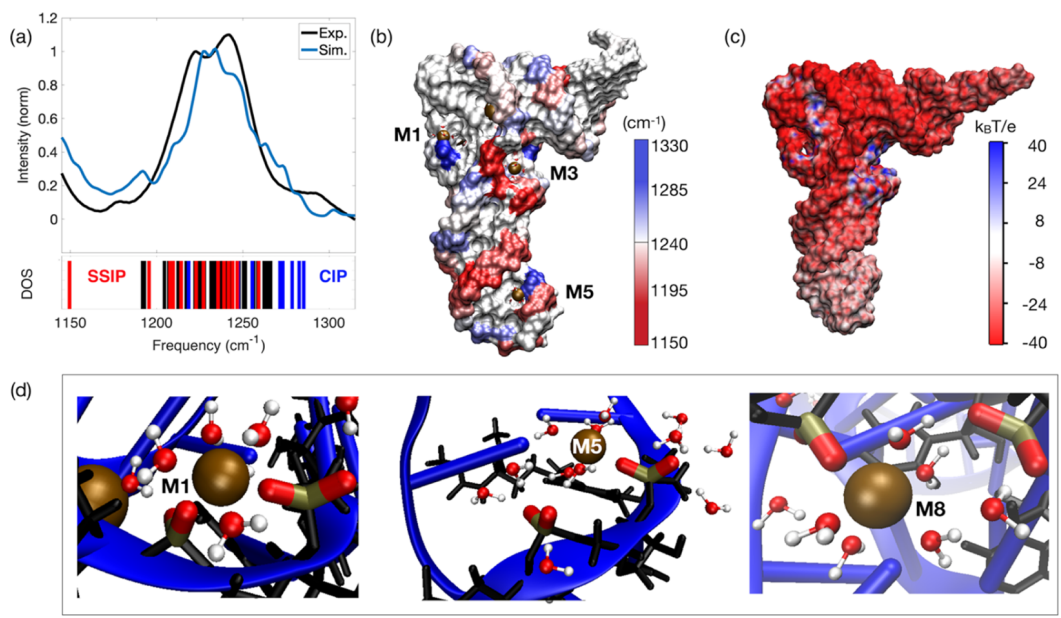

**Figure 4.** Results of ab-initio QM/MM and MD simulations for tRNA$^{Phe}$ from yeast. (a) Simulated and experimental linear infrared absorption spectra in the frequency range of the asymmetric phosphate stretching vibration $\nu_{AS}(PO_2)^-$. Simulations are compared to the experimental infrared spectrum of undialyzed tRNA$^{Phe}$ in water. The simulated vibrational DOS of the $\nu_{AS}(PO_2)^-$ band color-codes the frequency positions of contact ion pairs of the $(PO_2)^-$ group with $Mg^{2+}$ ions (CIP, blue) and of SSIPs (red). Frequency positions of the remaining $(PO_2)^-$ groups are indicated in black. (b) Simulated spatial distribution of $\nu_{AS}(PO_2)^-$ vibrational frequencies. The surface color of the sugar–phosphate backbone encodes the local $\nu_{AS}(PO_2)^-$ frequency. (c) Electrostatic surface potential evaluated for 3200 snapshots at the end of a 1 $\mu$s MD trajectory. (d) Prototype solvation geometries around $(PO_2)^-$ groups with the first-solvation shell water molecules around $(PO_2)^-$ groups and $Mg^{2+}$ ions shown in the ball and stick representation. Solvation structure M1 (left) shows bidentate inner-sphere coordination of $Mg^{2+}$ by two adjacent $(PO_2)^-$ groups ($\nu_{AS}(PO_2)^-$ = 1285 and 1247 cm$^{-1}$), M5 (middle) shows contact ion pair formation of the $(PO_2)^-$ group and $Mg^{2+}$ in the anticodon loop with the first-solvation shell waters shared by $Mg^{2+}$ and the adjacent $(PO_2)^-$ group ($\nu_{AS}(PO_2)^-$ = 1278 and 1208 cm$^{-1}$), M8 shows a CIP within the D-loop and SSIP mediated interstrand contact to the $T\Psi C$ loop ($\nu_{AS}(PO_2)^-$ = 1282 and 1210 cm$^{-1}$); $Mg^{2+}$ ions are shown in ochre, phosphorous atoms in dark yellow, oxygen atoms in red, and hydrogen atoms in white. Except for the $(PO_2)^-$ units, the tRNA backbone is shown in black.

The spatial mapping of $\nu_{AS}(PO_2)^-$ frequencies to the surface of tRNA$^{Phe}$ (Figure 4b) exhibits strong local variations and minor homogeneity for different domains of tRNA$^{Phe}$. The frequency positions $\nu_{AS}(PO_2)^-$ of particular $(PO_2)^-$ groups are determined by their local hydration geometry, which is found to involve mostly two adjacent phosphate groups. The restriction to two neighboring phosphate groups in the QM region of QM/MM simulations inherently assumes local hydration structures that span single- to diphosphate-ribose segments. For exceptions where nearby phosphate groups approach each other in crowded regions of tRNA$^{Phe}$, we have verified the accuracy of the approach in benchmark simulations covering up to four phosphate groups (data not shown). Because of the high water accessibility[27] in the helical domain of the anticodon of tRNA$^{Phe}$, this region is exceptional with a more homogeneous frequency distribution $\nu_{AS}(PO_2)^-$ ∼ 1220 cm$^{-1}$.

Figure 4d shows prototypical local hydration geometries at different $(PO_2)^-$ sites. We observe a pronounced blue-shift of $\nu_{AS}(PO_2)^-$ for bidentate inner-sphere complexation of $Mg^{2+}$ ions by two $(PO_2)^-$ groups (M1 in Figure 4d, $\nu_{AS}(PO_2)^-$ = 1285 and 1247 cm$^{-1}$ of the two $(PO_2)^-$ units). Singly coordinated CIPs show blue-shifted $\nu_{AS}(PO_2)^-$ frequencies in the range ∼1250–1280 cm$^{-1}$ (M5 in Figure 4d, $\nu_{AS}(PO_2)^-$ = 1278 cm$^{-1}$). Here, water molecules in the first solvation shell of the $Mg^{2+}$ ion form hydrogen bonds and are part of the hydration shell of the adjacent $(PO_2)^-$ group. For such SSIP configurations adjacent to a CIP, we find a red-shift of $\nu_{AS}(PO_2)^-$ to 1200–1220 cm$^{-1}$ (Figure 4d, SSIP with M5: $\nu_{AS}(PO_2)^-$ = 1208 cm$^{-1}$), representing a substantial >60 cm$^{-1}$ spread of the asymmetric stretching frequencies on a sub-5 Å length scale. The red-shift of $\nu_{AS}(PO_2)^-$ is caused by an ion-induced ordering of the water arrangement around the $(PO_2)^-$ group in the SSIP by which the local electric field acting on the $(PO_2)^-$ group is enhanced. This result correlates with experimental findings of increased absorption in the range 1180–1230 cm$^{-1}$ upon addition of small amounts of $Mg^{2+}$ ions to dialyzed tRNA$^{Phe}$ ($R$ = 4.8, cf. Figure 2b). Similar local hydration geometries are found for M6 and M4. For the latter, a magnesium-containing SSIP bridges the deep and narrow groove in a helical region of tRNA$^{Phe}$. At position M8, contact between the D-loop and the $T\Psi C$ loop is mediated via a SSIP configuration that induces ordered water molecules (Figure 4d, CIP: $\nu_{AS}(PO_2)^-$ = 1282 cm$^{-1}$; SSIP: $\nu_{AS}(PO_2)^-$ = 1210 cm$^{-1}$).

We have further analyzed the effective electrostatic potential at the surface of tRNA$^{Phe}$ (Figure 4c). For the helical domains of the acceptor stem and the anticodon region, we find the typical negative surface potential on the order of $-40k_BT/e \approx -1.0$ V ($k_BT$: thermal energy at a temperature $T = 298$ K, $e$: elementary charge), due to the negatively charged $(PO_2)^-$ groups and in qualitative agreement with findings for double-stranded RNA.[7] However, in the crowded regions of tRNA$^{Phe}$ (D and T$\Psi$C loop), the negative electrostatic potential due to the high charge density of $(PO_2)^-$ oxygens is fully compensated by the presence of a small number of immobilized (contact) $Mg^{2+}$ ions, locally inducing a net positive effective surface potential (cf. Figure 1a: M1, M3, M7, M2, M8). The contact interactions of $Mg^{2+}$ ions with $(PO_2)^-$ groups thus (over)compensate the repulsive Coulomb interaction and stabilize the tertiary structure of tRNA. Similarly, the low electrostatic surface potential in the anticodon region arises from the compensation of negative $(PO_2)^-$ charges in the presence of the $Mg^{2+}$ ion together with particular high solvent accessibility (Figures 1a and 4c,d: position M5).

## 4. DISCUSSION

The combination of dialysis and linear infrared spectroscopy gives insights into interaction patterns between $Mg^{2+}$ ions and phosphate groups in the backbone of tRNA. Starting from tRNA$^{Phe}$ and E.c. tRNA samples with negligible magnesium content, the number of interacting $Mg^{2+}$ ions rises linearly with the concentration ratio $R = c(Mg^{2+})/c(tRNA)$, as shown in Figure 1b. Up to $R \approx 7$, all added $Mg^{2+}$ ions interact with the tRNAs. At higher $Mg^{2+}$ concentrations, only a fraction of ions interacts with the tRNAs, leading to the deviation from a linear behavior in Figure 1b. There are no indications of cooperativity of the $Mg^{2+}$ uptake in the concentration shown in Figure 1, a conclusion in line with previous dialysis studies at lower tRNA concentrations.[25]

The linear infrared absorption spectra of tRNA$^{Phe}$ (Figure 2) exhibit two strong components with maxima at 1220 and 1241 cm$^{-1}$ and the shoulder at 1270 cm$^{-1}$. The component around 1220 cm$^{-1}$ is due to phosphate groups fully exposed to water with separate hydration shells consisting of up to 6 water molecules and a prototypical tetrahedral hydrogen-bond arrangement around the $(PO_2)^-$ oxygens.[27] The absorption around 1241 cm$^{-1}$ is due to $\nu_{AS}(PO_2)^-$ vibrations of phosphate groups with an under-coordination in the number of water molecules, including "ordered" hydration environments consisting of chain-like arrangements of water molecules. The absorption around 1270 cm$^{-1}$, a prominent component of the differential absorption spectrum of Figure 2b, is a hallmark of CIP formation, as is evident from the theoretical calculations and previous work on model systems.[17,18]

The CIP infrared absorption around 1270 cm$^{-1}$ rises with the $Mg^{2+}$ concentration (Figure 2b). Its peak value saturates as a function of the concentration ratio $R$ (Figure 1c) but at much lower $Mg^{2+}$ concentrations than the number of $Mg^{2+}$ ions interacting with tRNA$^{Phe}$ and E.c. tRNA (Figure 1b). The differential absorbance at 1270 cm$^{-1}$ (Figure 1c) reaches a value of up to 3% of the peak absorbance of tRNA at 1240 cm$^{-1}$. Assuming a similar molar extinction coefficient of the $\nu_{AS}(PO_2)^-$ vibrations of CIPs and phosphate groups without a $Mg^{2+}$ ion nearby, one estimates a minimum number of 3 CIPs per tRNA molecule. On the other hand, the relative strengths of the 2D-IR signals at 1240 and 1270 cm$^{-1}$ (cf. Supporting Information, Table S1) suggest the existence of $6 \pm 2$ CIPs per tRNA for $R = 15$. We consider the latter number an upper limit of the number of CIPs per tRNA. The CIPs are expected to be formed at sites with a high negative charge density from phosphate groups, like at the sites M3, M7/M8, M1, and M2 (Figures 1a and 4d). The discrete number of $Mg^{2+}$ ions inverts the sign of the effective electrostatic surface potential, thus stabilizing the tertiary tRNA structures locally.

Our experimental and theoretical results provide clear evidence for the existence of CIPs in the equilibrium structures of tRNA$^{Phe}$ and E.c. tRNA. Such CIPs represent the "strongly interacting ion species", which has been discussed in the literature.[25] Their impact on the electrostatic potential at the crowded sites of tRNA (M1−M2, M3, M7−M8) is much stronger than the contribution from long-range electric fields originating from the distant outer ion atmosphere and from contact ion pairs with Na$^+$ (cf. Supporting Information). This fact shows that CIPs play a prominent role for stabilizing the tertiary folded cloverleaf structure of tRNA. It should be noted that the water molecules around phosphate groups without $Mg^{2+}$ ions make a major contribution to the electrostatic potential (cf. Figure S11b).

Our results are in contrast to predictions from PB treatments, claiming that $Mg^{2+}$ ions solvated in the outer ion atmosphere were the structure-stabilizing constituents.[24] The surface electrostatic potentials derived in ref 24 are substantially lower ($\sim$20%) than the potentials shown in Figure 4c. Such small potentials fail to account for the electric field-dependent frequency positions of the $\nu_{AS}(PO_2)^-$ vibrations.[28,29] PB theory neglects the direct contribution of water molecules to the electrostatic potential and uses the static dielectric constant of water to scale the bare Coulomb interaction potential. Given the subtle balance of attractive and repulsive molecular interactions in this complex many-body system of fluctuating charges, such two approximations appear inappropriate.

The 2D-IR spectra presented in Figure 3 give information on dynamics at the molecular scale and on interactions between the different charged and polar constituents of hydrated tRNA. The 2D-IR spectra display strong overlapping diagonal peaks (yellow-red contours) around detection frequencies $\nu_3 = 1220$ and 1245 cm$^{-1}$, which are complemented by a shoulder-like feature around 1270 cm$^{-1}$, the strength of which rises with the $Mg^{2+}$ ion concentration. There are no cross-peaks in any of the 2D-IR spectra, that is, vibrational couplings between the different diagonal components are minor. This fact is a clear indication that the different diagonal contributions originate from phosphate groups, which are mainly uncoupled and embedded in different local environments.

For a quantitative analysis of the line shapes in the 2D-IR spectra of tRNA$^{Phe}$, we performed simulations based on a density matrix approach for describing the nonlinear vibrational response.[30] This treatment includes four vibrational transitions centered at 1220, 1245, 1270, and 1280 cm$^{-1}$ (cf. Figure 2). The frequency fluctuation correlation function (FFCF) of the aqueous environment is accounted for by a Kubo ansatz with two exponential terms of 300 fs and 50 ps decay time. The simulated line shapes include lifetime broadenings which are calculated with vibrational lifetimes of 290 fs for the 1220 and 1245 cm$^{-1}$ components and 700 fs for the 1270 and 1280 cm$^{-1}$ contributions. A comparison of experimental and simulated spectra is presented in Supporting Information (Figures S4 and S5) and shows good agreement in the overall line shapes.

Of particular interest is the 2D-IR signal around $\nu_3 = 1270$ cm$^{-1}$ which is due to CIPs and much more pronounced than the linear absorption at 1270 cm$^{-1}$ in the $\nu(PO_2)^-$ absorption spectrum (Figure 2). The higher relative amplitude in the 2D-IR spectra is mainly caused by (i) the longer vibrational lifetime of the 1270 cm$^{-1}$ excitations in comparison to those at 1220 and 1245 cm$^{-1}$ (700 vs 290 fs) and (ii) reduced amplitude of the fast fluctuation component in the FFCF. At a population time $T = 300$ fs at which the 2D-IR spectra of Figure 3 were recorded, the 1220 and 1245 cm$^{-1}$ signals have decayed to some 35% of their maximum value, while the 1270 cm$^{-1}$ contribution is at 70% of its initial value. The reduced amplitude of the fast decay in the FFCF points to a more rigid hydration structure in CIP environments, partly due to the strong impact of the strong local electric fields on the orientation of water molecules. Such experimental observations are corroborated by results from MD simulations that analyze the tRNA$^{Phe}$ first-solvation shell. The hydrogen bond angular distribution of hydration geometries in the first solvation shell around phosphate groups (Figure S10) shows a bimodal distribution arising from direct hydrogen bonds with the oxygen atoms of the phosphate group and water molecules that occupy the first solvation shell of Mg$^{2+}$ ions in CIPs. The predominant contribution to the hydration geometries of water molecules in the first solvation shell of Mg$^{2+}$ ions is characterized by vibrational frequencies $\nu(PO_2)^- > 1255$ cm$^{-1}$ in the simulated linear infrared absorption spectra (Figures 4a and S9). The results are, thus, indicative of a more rigid hydration shell around CIPs, as manifested in the smaller fluctuation amplitudes in 2D spectra, where a substantially smaller width of antidiagonal 2D-cuts at 1270 cm$^{-1}$ is observed compared to 1245 cm$^{-1}$.

## 5. CONCLUSIONS

In conclusion, we have studied the electrostatic properties of tRNA$^{Phe}$ and E.c. tRNA embedded in an aqueous environment, which contains Mg$^{2+}$ ions. A combination of dialysis, fluorescence spectroscopy, linear infrared spectroscopy, femtosecond 2D-IR spectroscopy, MD simulations, and ab-initio calculations of the tRNA$^{Phe}$ sugar–phosphate vibrational frequencies gives evidence of a prominent role of Mg$^{2+}$–phosphate contact ion pairs in stabilizing the folded tertiary structure of tRNA. The formation of contact ion pairs is manifested in a blue-shift of the infrared transition of the asymmetric $(PO_2)^-$ stretching vibration to frequencies around 1270 cm$^{-1}$, a behavior present in both the linear and the 2D-IR spectra. Up to six contact ion pairs are formed per tRNA, predominantly at positions with a high negative charge density from phosphate groups. Addition of a Mg$^{2+}$ ion to such distinguished sites results in stabilization of the folded tertiary structure because of strong attractive electrostatic interactions. The Mg$^{2+}$ contact sites found in the present work agree with results from X-ray diffraction studies. The double-helical parts of the folded tRNA structures display an overall negative surface potential which is compensated for by water molecules as well as Mg$^{2+}$ and other cations separated by one or more water layers from tRNA. Our results underline the need for probing electric fields at the local molecular level and for atomistic simulations of the local interactions geometries. They demonstrate the predominance of local over long-range electrostatic interactions in defining the tertiary RNA structure.

## ■ ASSOCIATED CONTENT

### sı Supporting Information

The Supporting Information is available free of charge at https://pubs.acs.org/doi/10.1021/acs.jpcb.0c08966.

Materials and methods, results of pump−probe experiments, simulation of 2D-IR spectra, and theory results (PDF)

## ■ AUTHOR INFORMATION

### Corresponding Author

Thomas Elsaesser − *Max-Born-Institut für Nichtlineare Optik und Kurzzeitspektroskopie, Berlin 12489, Germany*; orcid.org/0000-0003-3056-6665; Email: elsasser@mbi-berlin.de

### Authors

Jakob Schauss − *Max-Born-Institut für Nichtlineare Optik und Kurzzeitspektroskopie, Berlin 12489, Germany*

Achintya Kundu − *Max-Born-Institut für Nichtlineare Optik und Kurzzeitspektroskopie, Berlin 12489, Germany*; orcid.org/0000-0002-6252-1763

Benjamin P. Fingerhut − *Max-Born-Institut für Nichtlineare Optik und Kurzzeitspektroskopie, Berlin 12489, Germany*; orcid.org/0000-0002-8532-6899

Complete contact information is available at:
https://pubs.acs.org/10.1021/acs.jpcb.0c08966

### Author Contributions

J.S. and A.K. contributed equally to this research. T.E. and B.P.F. initiated, conceived and supervised the study. J.S. and A.K. performed the experiments, B.P.F. did the MD calculations. All authors contributed to the analysis of the results, T.E. and B.P.F. wrote the manuscript with input from all authors.

### Notes

The authors declare no competing financial interest.

## ■ ACKNOWLEDGMENTS

This research has received funding from the European Research Council (ERC) under the European Union's Horizon 2020 research and innovation program (grant agreements nos. 833365 and 802817). B.P.F. acknowledges support by the DFG within the Emmy-Noether Program (grant no. FI 2034/1-1). We thank Janett Feickert for technical support.

## ■ REFERENCES